\documentclass[prl, twocolumn,superscriptaddress,nobibnotes]{revtex4-2}
\usepackage{graphicx}
\usepackage{amssymb}
\usepackage{amsmath}
\usepackage{bm}
\usepackage{color}
\usepackage{float}
\usepackage{setspace}
\usepackage{physics}
\usepackage{subfigure}

\usepackage{lineno}

\usepackage[T1]{fontenc}
\usepackage{ulem}

\begin{document}
	\title{Splitting spin-orbit coupled polariton vortex pairs
		\\ in the non-Hermitian Rashba-Dresselhaus band at room temperature}

	\author{Xiaokun Zhai}
	\affiliation{Department of Physics, School of Science, Tianjin University, Tianjin 300072, China}

	\author{Ying Gao}
	\affiliation{Department of Physics, School of Science, Tianjin University, Tianjin 300072, China} 
	
	\author{Xuekai Ma}
	\affiliation{Department of Physics and Center for Optoelectronics and Photonics Paderborn (CeOPP), Universit\"{a}t Paderborn, Warburger Strasse 100, 33098 Paderborn, Germany}
	
	\author{Chunzi Xing}
	\affiliation{Department of Physics, School of Science, Tianjin University, Tianjin 300072, China} 
	
	\author{Xinmiao Yang}
	\affiliation{Department of Physics, School of Science, Tianjin University, Tianjin 300072, China} 
	
	\author{Xiao Wang}
	\affiliation{College of Materials Science and Engineering, Hunan University, Changsha 410082, China}
	
	\author{Anlian Pan}
	\affiliation{College of Materials Science and Engineering, Hunan University, Changsha 410082, China}
	
	\author{Marc Assmann}
	\affiliation{Department of Physics, Otto-Hahn-Str. 4a, TU Dortmund University, 44227 Dortmund, Germany}
	
	\author{Stefan Schumacher}
	\affiliation{Department of Physics and Center for Optoelectronics and Photonics Paderborn (CeOPP), Universit\"{a}t Paderborn, Warburger Strasse 100, 33098 Paderborn, Germany}
	\affiliation{Institute for Photonic Quantum Systems (PhoQS), Paderborn University, 33098 Paderborn, Germany}
	\affiliation{Wyant College of Optical Sciences, University of Arizona, Tucson, AZ 85721, USA}
	
	\author{Tingge Gao}
	\affiliation{Department of Physics, School of Science, Tianjin University, Tianjin 300072, China}

	\begin{abstract}
		{Spin orbit coupling gives rise to intriguing physical phenomena in bosonic condensates, such as formation of stripe phases and domains with vortex arrays. However, how the non-Hermiticity affects the spatial distribution of spin orbit coupled topological defects such as vortex pair is still challenging to study. In the present work, we realize a non-equilibrium room-temperature exciton polariton condensate within a microdisk potential in a liquid crystal (LC) microcavity with the perovskite CsPbBr$_3$ as optically active material. We use the interplay of TE-TM mode splitting and Rashba-Dresselhaus spin-orbit coupling (RDSOC) to realize electrically tunable polariton vortex pairs with locked spin and orbital angular momentum. Importantly, the non-Hermiticity of RDSOC bands leads to nonreciprocal transportation of the vortex pair such that they move to the opposite edges of the microdisk depending on their spin. Our results are robust against sample imperfections and pave the way to investigate the coupling of vortex orbital and spin degrees of freedom in a quantum fluid of light at room temperature, offering potential for generation of complex states of light for non-Hermitian quantum optical information processing within optoelectronic chips.}

	\end{abstract}
	
	\maketitle
	Coupling of spin and orbital degrees of freedom of particles determines the intrinsic properties of many materials. Enabling the control of spin degrees of freedom in solids \cite{1-SOI review,2-SOI book}, spin-orbit coupling (SOC) plays a critical role in spintronics and topological insulators \cite{5-Kato Scienece, 6-Spinhalleffect2, 7-topological insulator1, 8-topological insulator2, 9-topological insulator3}. 
	In the regime of equal Rashba \cite{4-Rashba} and Dresselhaus \cite{3-Dresselhaus} spin-orbit coupling infinite spin lifetime \cite{Shoucheng}, spin helices and spin precession were demonstrated \cite{5-Kato Scienece, 6-Spinhalleffect2} in a two dimensional electron gas. In bosonic cold atom condensates, synthetic spin-orbit coupling is explored, which can be controlled by careful engineering of the spin states of atoms with laser beams. There, for example the transition from a spin-mixed phase to a spin-separated phase \cite{10-SOI atom BEC} and the measurement of a stripe phase \cite{Wolfgang Ketterle} in a one dimensional Rashba and Dresselhaus coupled cold atom gas have been realized. By tuning the Raman coupling strength and introducing spatially dependent detuning which creates an effective magnetic field in the RDSOC regime \cite{RD theory1, RD theory2, RD review}, vortices or vortex lattices can be formed \cite{Zhai hui, spielman vortex1, spielman vortex2}. In these systems, complicated arrangement of multiple lasers is needed to realize the underlying Hamiltonian with effective SOC.

	\begin{figure}[t]
		\centering
		\includegraphics[width=\linewidth]{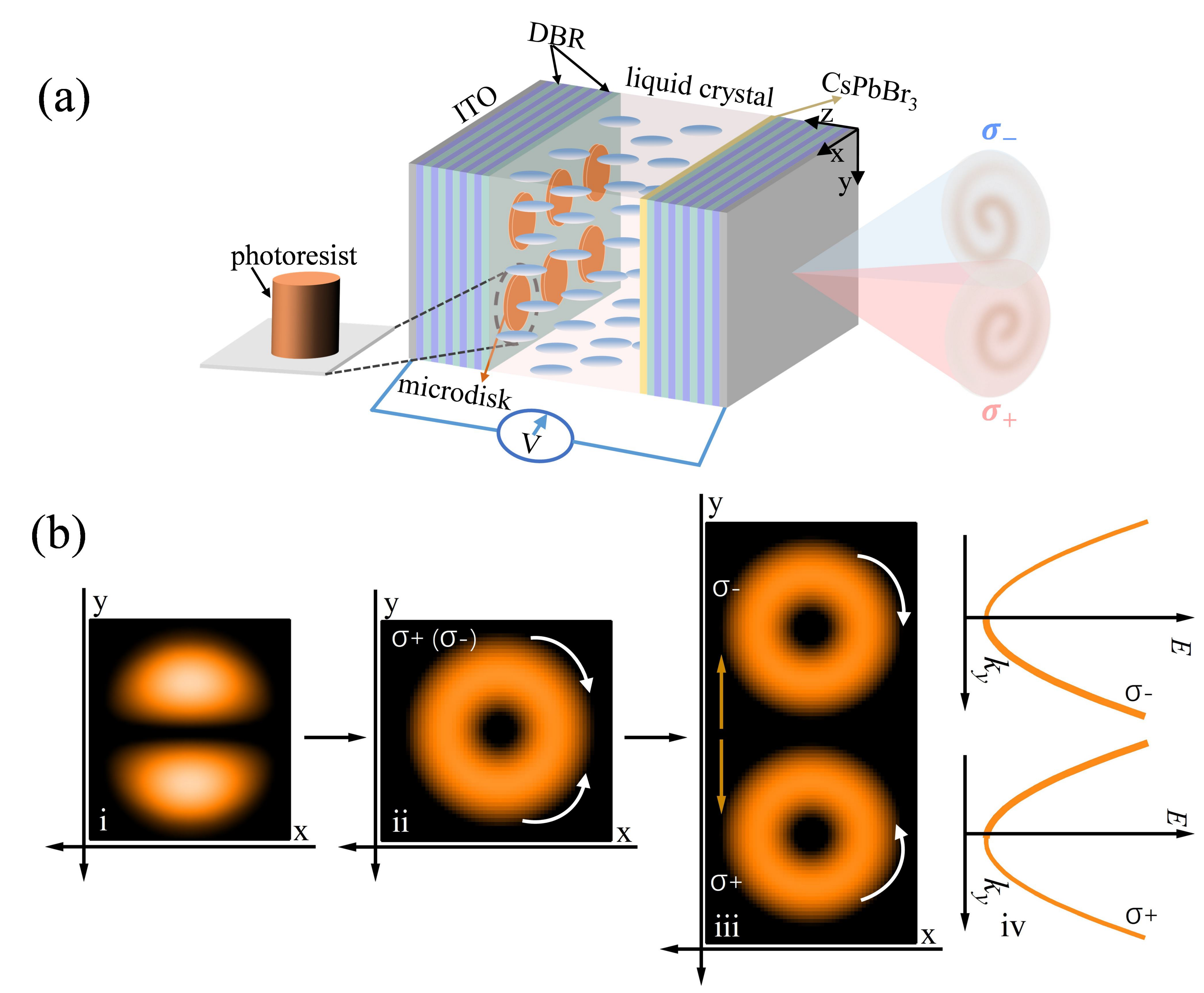}
		\caption{\textbf{Schematic of the LC microcavity and formation of spin-orbit coupled vortex pairs.} (a) The microcavity is filled with LC and contains a microdisk array and CsPbBr$_3$ microplates facing each other. (b) Sketch of the idealized modes in the microcavity: the TE-TM splitting splits a polariton dipole mode into two spin-orbit-locked vortices spatially separated by the non-Hermiticity of the RDSOC bands.}
	\end{figure}
	
	In this work we demonstrate the exciton polariton vortex pairs with locked spin and orbital degrees of freedom that form as robust topological objects in the Rashba-Dresselhaus regime in a photonic planar microresonator at room temperature. In such optical systems vortices formed with well defined and locked spin and orbital degrees of freedom play an increasingly important role for communication technologies and qubit manipulation schemes \cite{OAM application, topological vortex laser1, topological vortex laser2}. As depicted in Figure~1(a), our realization constitutes a LC planar photonic microcavity. The LC allows to tune the spatial anisotropy and related effective refractive indices with an externally applied electric field, such that an effective RDSOC \cite{1-liquid crystal_science} is realized with a Hamiltonian of the form $H=\hbar^2 k^2/2m+2\alpha k_y\sigma_z$. $\alpha$ represents the RDSOC strength that leads to the separation of the two opposite spin components \cite{LC PRL}($\sigma_z$ is the Pauli matrix).

	The particles underlying our system are hybrid quasiparticles, so-called exciton polaritons, that are created due to the strong coupling of excitons and cavity photons \cite{polariton BEC1, polariton BEC2}. These have attracted significant attention for the manipulation of vortex states \cite{lagoudakis1, resonant vortex 2, lagoudakis rotate vortex, Ma xuekai,Fraser, BIC vortex, Deveaud} for instance and enabled by the large exciton binding energy and oscillator strength in the perovskites show condensation at room temperature \cite{Yao RD, Xiaokun vortex, gao ying, zhuxiaoyang, 21-xiong condensation}. A splitting between longitudinal and transversal optical modes (or TE-TM splitting) is intrinsic to such planar microcavities and acts as an effective magnetic field \cite{non-Abelian gauge field}, giving rise to the optical spin Hall effect \cite{kavokin spin hall effect, spin hall effect nature physics, nonlinear optical spin hall effect}. In addition, the TE-TM splitting influences the polariton modes formed: considering a setting with near rotational symmetry and spatially confined topological defects with finite orbital angular momenta, a dipole mode ((i) in Figure 1(b)) is split into two vortices with topological charges locked with the vectorial polarization degree of freedom ((ii) in Figure 1(b)), or in other words, locked with the SAM \cite{TETM vortex, PRL polarization vortex}. In these vortex pairs where spin and orbital degrees of freedom are locked, the vortex that is observed in one circularly polarized component (e.g., $\sigma-$) rotates in the opposite direction (e.g., clockwise) from the vortex observed in the other circularly polarized component ($\sigma+$; counter-clockwise rotation). This can be described by a 2$\times$2 matrix \cite{Yao RD} in the circular polarization basis: 
	\begin{equation}
	H({\bf{k}})=
	\begin{bmatrix}
	\dfrac{\hbar^2{\bf{k^2}}}{2m} & \beta{\bf{k^2}}e^{2i\varphi} \\
	\beta{\bf{k^2}}e^{-2i\varphi} & \dfrac{\hbar^2{\bf{k^2}}}{2m} \\
	\end{bmatrix}.
	\end{equation}
	Here, $\beta$ represents the strength of the TE-TM splitting.  The underlying modes can be obtained by numerically solving the Schr{\"o}dinger equation $[H+V]\Psi=E\Psi$, with confinement potential $V$ in the circular shape as illustrated in the middle graph (ii) of Figure~1(b) and discussed in more detail in the SM. However, these kinds of well defined spin-orbit locked modes that also form a natural basis for the interpretation of the condensate modes have not been observed yet at room temperature. With only weak TE-TM splitting, a very high-quality microcavity would be required as otherwise detrimental factors such as sample disorder or interaction between polaritons tend to destroy the locking between SAM and OAM.

	\begin{figure*}[t]
		\centering
		\includegraphics[width=0.8\linewidth]{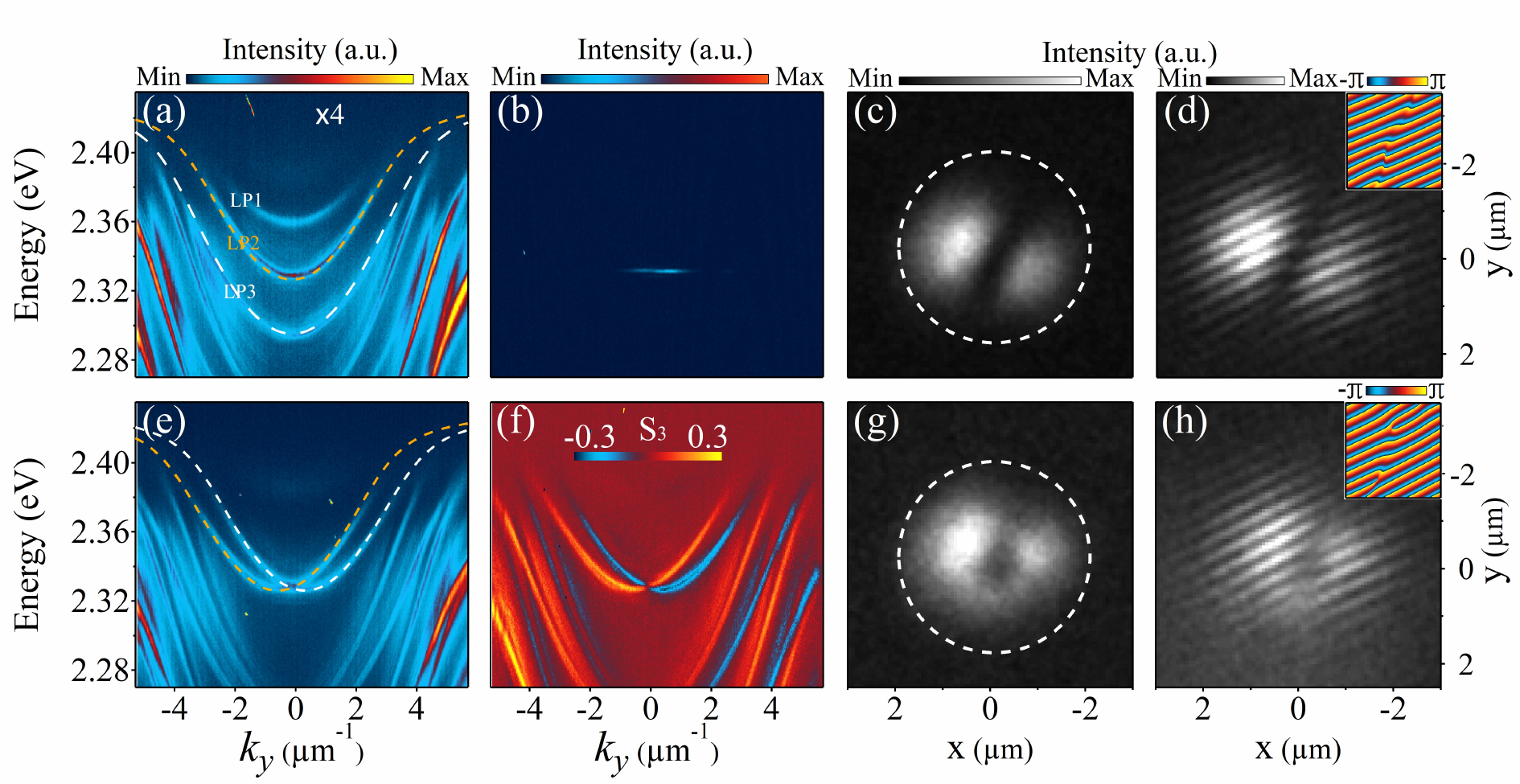}
		\caption{\textbf{Polariton condensation in the LC microcavity at different voltages.} (a, b) Dispersion below and above threshold at 5 V. (c, d) Real space image and interferogram of the polariton condensate at 5 V. (e) Dispersion below threshold at 5.6 V. (f) Spin polarized (S$_3$ component) dispersion at 5.6 V, defined as ($I_\textup{left}$-$I_\textup{right}$)/($I_\textup{left}$+$I_\textup{right}$) where $I_\textup{left}$ ($I_\textup{right}$) is the left(right)-hand circularly polarized dispersion. Fitted curves are the simulated bands using a coupled oscillator model (a) and a four-band Hamiltonian \cite{Yao RD} (e). (g, h) Real space image and interferogram of the polariton condensate at 5.6 V. The dashed lines in (c, g) indicate the microdisk. The insets in (d, h) show the extracted phase. }
	\end{figure*}
	
	The polarization dependent loss \cite{polarization loss} of the horizontal and vertical linearly polarized polariton modes can lead to asymmetric linewidth when they are brought into resonance to form non-Hermitian RDSOC bands, which indicates nonreciprocal transport and leads to the spin dependent spatial separation ((iii) and (iv) in Figure 1(b)) \cite{skin effect1, skin effect2, skin effect3, skin effect4, skin effect5}. In the present work, we show the two spin orbit coupled vortices created under TE-TM splitting are spatially separated through the non-Hermicity induced nonreciprocal transport of the RDSOC bands, such that each spin component of the polariton condensate carries a distinct OAM and moves towards the opposite edge of the potential trap. The formation of spin-orbit locked vortex pairs can be directly observed by measuring the left- and right-circularly polarized emission at room temperature. These findings are robust against disorder within the microcavity. Our results demonstrate a different mechanism to electrically create and separate vortex with locked spin and orbital degrees of freedom by the non-Hermitian transportation in a quantum fluid of light, offers to investigate non-Hermitian skin effect in the spin orbit coupled bosonic condensate in the future.

	In the experiment we introduce photoresist microdisk structures with the diameter of 4 $\mu$m (the distance between the microdisk is around 15 $\mu$m to avoid interaction) and the height of 120 nm onto the bottom Distributed Bragg Reflector (DBR) using lithography technique in the LC microcavity, where the effective cavity length of the microdisk is larger and the cavity mode energy is smaller compared with other area. The details of the microcavity is shown in Figure 1(a) and the method section. In this case, the microdisks act as potential traps and polaritons are confined in these structures, occupying the specific discrete energy levels \cite{mesa GaAs Bloch1, mesa GaAs Bloch2}, for example, a dipole mode (see Figure 1(b)). The size of the inserted CsPbBr$_3$ microplates is around 40 $\mu$m which can cover at least one microdisk potential. We use a linearly polarized laser (repetition rate: 1 kHz, wavelength: 400 nm, pulse width: 50 fs) to excite the microcavity with the size of around 50 $\mu$m.

	We measure the dispersion of the LC microcavity as the same as \cite{Xiaokun vortex} below threshold when the voltage is 5 V, which is plotted in Figure 2(a). Multiple polariton branches are observed when the pumping density is far below the threshold due to the large thickness of the cavity. That is, several cavity modes exist within the microcavity and strongly couple with the excitons (detailed fitting parameters are shown in the SM). The strong coupling is confirmed at the large-wavevector region where one can clearly observe that the dispersions become flat when approaching the exciton resonance. Under this voltage, the LP1 and LP3 are horizontally linearly polarized whereas the LP2 and LP4 are vertically linearly polarized.

	
	\begin{figure*}[t]
		\centering
		\includegraphics[width=0.8\linewidth]{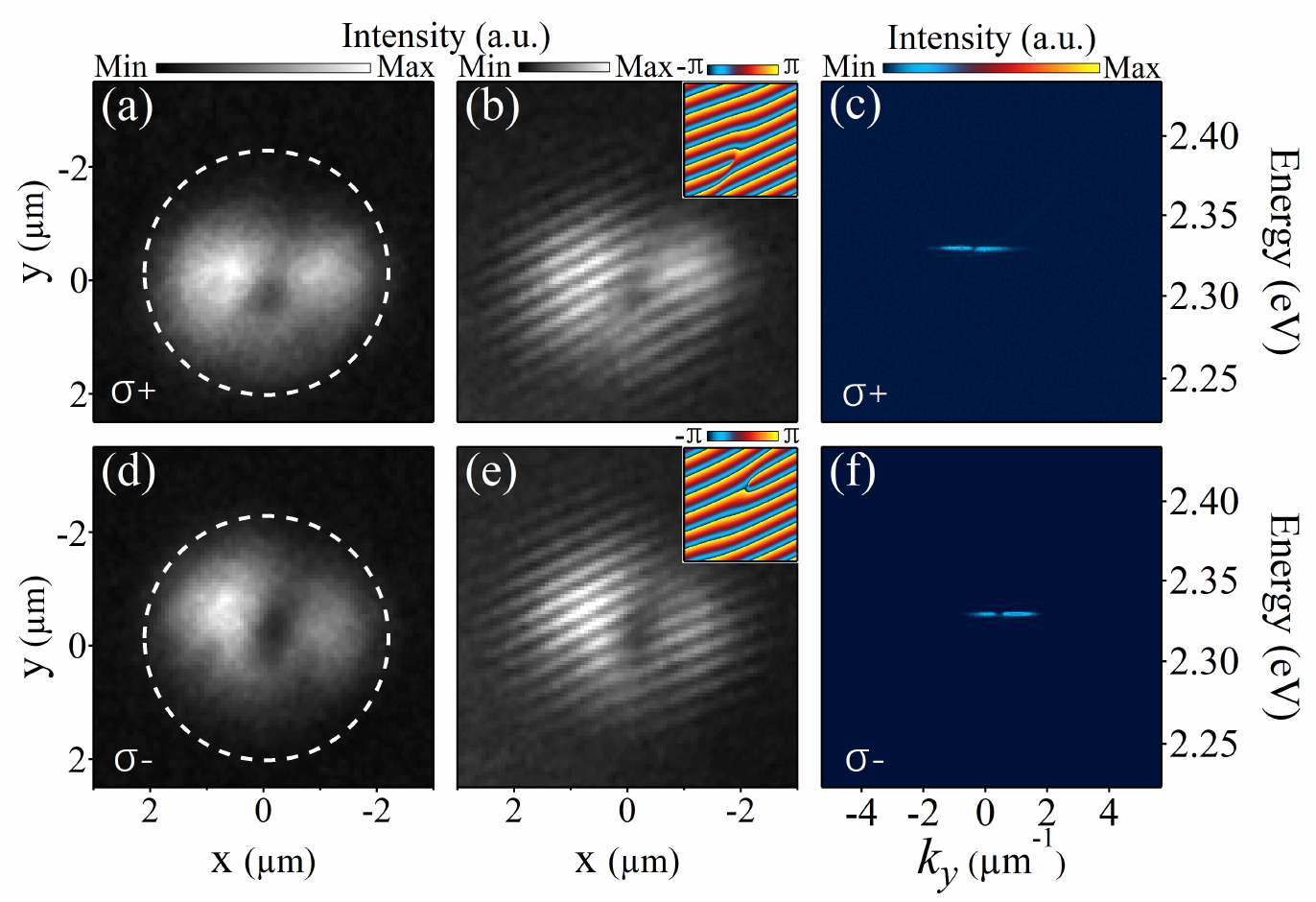}
		\caption{\textbf{Spin-orbit locked polariton vortex pair at 5.6 V.} (a, b, c) Real space image, interferogram and dispersion of the $\sigma+$ component at 5.6 V. (d, e, f) Real space image, interferogram and dispersion of the $\sigma-$ component at 5.6 V. The dashed lines in (a) and (d) indicate the microdisk. The insets in (b) and (e) are the extracted phase indicating the vortex.}
	\end{figure*}

	With further increasing the pumping density to around 18 $\mu$J/cm$^2$, the emitted PL intensity of the polaritons increases superlinearly, whereas the linewidth drops suddenly and the polariton energy shows noticeable continuous blueshift (details in SM). These results clearly show the occurrence of the polariton condensation at the lower branch LP2 (Figure 2(b)). We note that the thresholds of the modes in the microdisk potential can vary in a certain range, depending on the external pumping, relaxation and decay of polaritons \cite{amo prx, relax1, relax2}. In the experiments, we carefully choose the microdisk size, thickness and pumping such that the condensed polaritons form in a single dipole mode above the threshold in real space with orientation determined by the anisotropy of the potential, see Figure 2(c) (The degeneracy with another dipole mode is broken by the anisotropy thus does not affect the experimental results). We set the pumping density to be around 1.2 P$_{th}$ where the nonlinearity can be neglected. We build a Michelson interferometer where one arm is expanded by around 15 times and acts as the reference beam. Clear interference fringes are observed above the threshold (Figure 2(d)), the extracted phase from the Fast Fourier transformation of the fringes shown in the upper right corner indicates the development of the macroscopic coherence across the condensate and the phase jump of $\pi$ across the two lobes.

	The TE-TM splitting is prominent in anisotropic perovskite or LC-based microcavities \cite{LC berry curvature, sanvetto tuning berry curvature, TE-TM LC}, which can decompose the above dipole mode into two vortices with the topological charge of $\pm1$, locked with the SAM. However, the disorder within perovskite microplates or photoresist microdisk can affect the formation of the vortices. In addition, the interaction between the polaritons can also destroy the locking between the SAM and OAM, thus preventing the observation of the spin-orbit locked vortices. As analyzed above, these two vortices can be directly measured in the RDSOC regime, which can be realized by simply increasing the voltage in our experiments. When the voltage is increased to 5.6 V, the polariton branch LP3 is blueshifted to be resonant with LP2. These two polariton branches have the opposite linear polarizations and parity, thus the RD spin split bands form (please see the total (Figure 2(e)) and spin polarized (Figure 2(f)) dispersion). Under the same pumping density of Figure 2(c), polaritons condense and the dipole mode becomes deformed under the same pumping density (shown in Figure 2(g)) where the polariton condensate carries a ring shape, indicating the formation of vortices. The interferogram of the polariton condensate in Figure 2(h) shows the emergence of two forks oriented in opposite directions. This confirms the existence of two vortices with the topological charge of $\pm1$ close to the core region of the ring-shaped condensate, which can be unambiguously proven by the extracted phase plotted at the upper right corner, calculated by using the same method as in Figure 2(d).
	
	\begin{figure*}[t]
		\centering
		\includegraphics[width=0.75\linewidth]{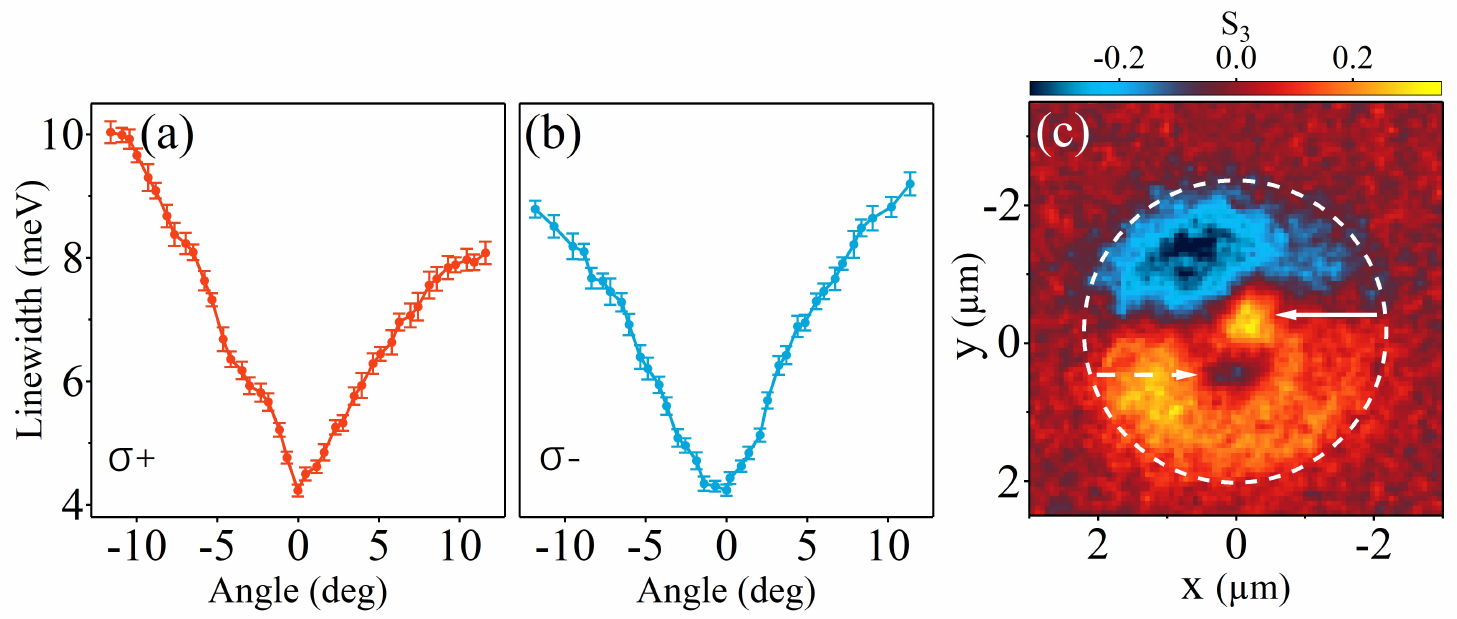}
		\caption{\textbf{Non-Hermiticity induced spatial separation of polariton vortex pair in the RDSOC bands.} (a) Linewidth of the polariton modes in the $\sigma+$ component at 5.6 V. (b) Linewidth of the polariton modes in the $\sigma-$ component at 5.6 V. Horizontal axis of (a) and (b) is shifted respectively to make the ground state of the two spin components to be at 0 for the clarity of the linewidth analysis. (c) Spin polarized real space images of the vortex pair. The arrows indicate the location of the two vortex cores. The dashed lines in (c) indicate the microdisk. }
	\end{figure*}
	
	To check whether the two vortices are locked with the definite SAM as we analyzed above, we measure the right- and left-hand circularly polarized PL emitted from the microcavity, which correspond to different spins states of the polariton condensate. In Figure 3(a), an intensity minimum in the center of the polariton condensate is clearly visible in the $\sigma+$ component, indicating the existence of a vortex. We note that the intensity distribution is not very symmetric around the intensity minimum, this is due to the inhomogeneity within the microcavity. From the interferogram by superimposing the real space image of the polariton condensate onto the reference arm, a fork is clearly observed, confirming the appearance of a vortex with the topological charge of -1, as plotted in Figure 3(b). The phase singularity can also be seen in the calculated phase distribution inserted at the upper right corner. From the dispersion taken under this voltage (Figure 3(c)), two peaks are observed to be located at the wavevector of 0.12 $\mu$m$^{-1}$ and -0.88 $\mu$m$^{-1}$, which is not symmetric along the normal incidence due to the RDSOC induced shift along $k_y$ direction. This kind of particular distribution of the dispersion confirms the appearance of the vortex with asymmetric distribution in the momentum space for the $\sigma+$ component. 
	
	On the other hand, in the $\sigma-$ component, both the intensity minimum in the real space image (Figure 3(d)) and the fork in the interferogram graph (Figure 3(e)) demonstrate the formation of the vortex with the topological charge of +1. The location of the polariton condensate is shifted upwards within the microdisk potential. In this spin component, the polariton condensate distribution in the momentum space is also shifted to the opposite direction compared to Figure 3(c), with the peak locations at 0.98 $\mu$m$^{-1}$ and -0.07 $\mu$m$^{-1}$ (Figure 3(f)). The mode distribution of the dispersion clearly confirms the existence of the swapped topological charge of the vortex in the $\sigma-$ component.

	The appearance of the spin-orbit locked polariton vortex pair under 5.6 V originates from the synthetic magnetic field due to the RDSOC and the prominent TE-TM splitting. The nonzero TE-TM splitting can be confirmed by measuring the two linearly polarized dispersions along $k_x$ direction when $k_y=0$ where the RDSOC disappears, shown in the SM. In our experiments, the TE-TM splitting decomposes the dipole mode into two spin-orbit locked vortices, whereas the spatial distribution of the two vortices is deeply affected by the non-Hermiticity of RDSOC bands. Experimentally, we calculate the linewidth of the polariton bands of the $\sigma+$ component below threshold based on the data in Figure 2(e) in the RDSOC regime, which show a clear asymmetric distribution along $k_y$ and -$k_y$ direction (Figure 4(a)). The linewidth of the polaritons along $k_y$ direction is smaller than -$k_y$ direction when their energy is the same, this means the vortex with the topological charge of -1 will nonreciprocally propagate along $k_y$ direction and accumulate at the bottom edge of the microdisk. For $\sigma-$ component, vortex with the topological charge of +1 will show opposite transportation behavior due to opposite asymmetry of the linewidth distribution (Figure 4(b)). Thanks to the spatially separation of the two vortices, the inter-species interaction can be reduced. In this case, the two vortices can be observed directly by measuring polarization resolved PL in the experiment. The spin polarized real space image of the polariton condensate at 5.6 V plotted in Figure 4(c) demonstrates the separation of the two vortices along \textit{y} direction with two cores clearly visible indicated by the arrows. We note that the separation of the polariton modes in the real space is different from \cite{LC PRL} where our experimental results are based on the non-Hermiticity of the RDSOC bands.

	If the RDSOC spin splitting disappears, for example, at 5 V where the polariton mode LP3 is far below LP2, the spin-orbit coupling vanishes. As a result, the two spin components of the polariton condensate share the same spatial mode distribution in the real space without definite vorticity and spin polarization (Figure 5(a-d)). When the voltage applied to the microcavity is increased to 6 V (such that LP3 is tuned above LP2; see the SM), the polariton condensate also assumes a dipole shape. The $\sigma+$ and $\sigma-$ spin-polarized components of the polariton condensate again show the same spatial mode without vorticity due to the absence of the SOC (Figure 5(e-h)). 
	
	Compared with other optically tunable spin-orbit coupled photonic vortex lasers \cite{topological vortex laser1, topological vortex laser2, Amo vortex} or polariton vortices confined in micropillar structures, which so far are limited to cryogenic temperature \cite{amo prx}, our work shows the non-Hermition RDSOC bands can nonreciprocally separate the spin orbit coupled vortex pair in the real space depending on the spin degree of freedom, and lays the foundation to manipulate electrically tunable state with spin-orbit coupled angular momentum at room temperature.
	
	\begin{figure}[t]
		\centering
		\includegraphics[width=\linewidth]{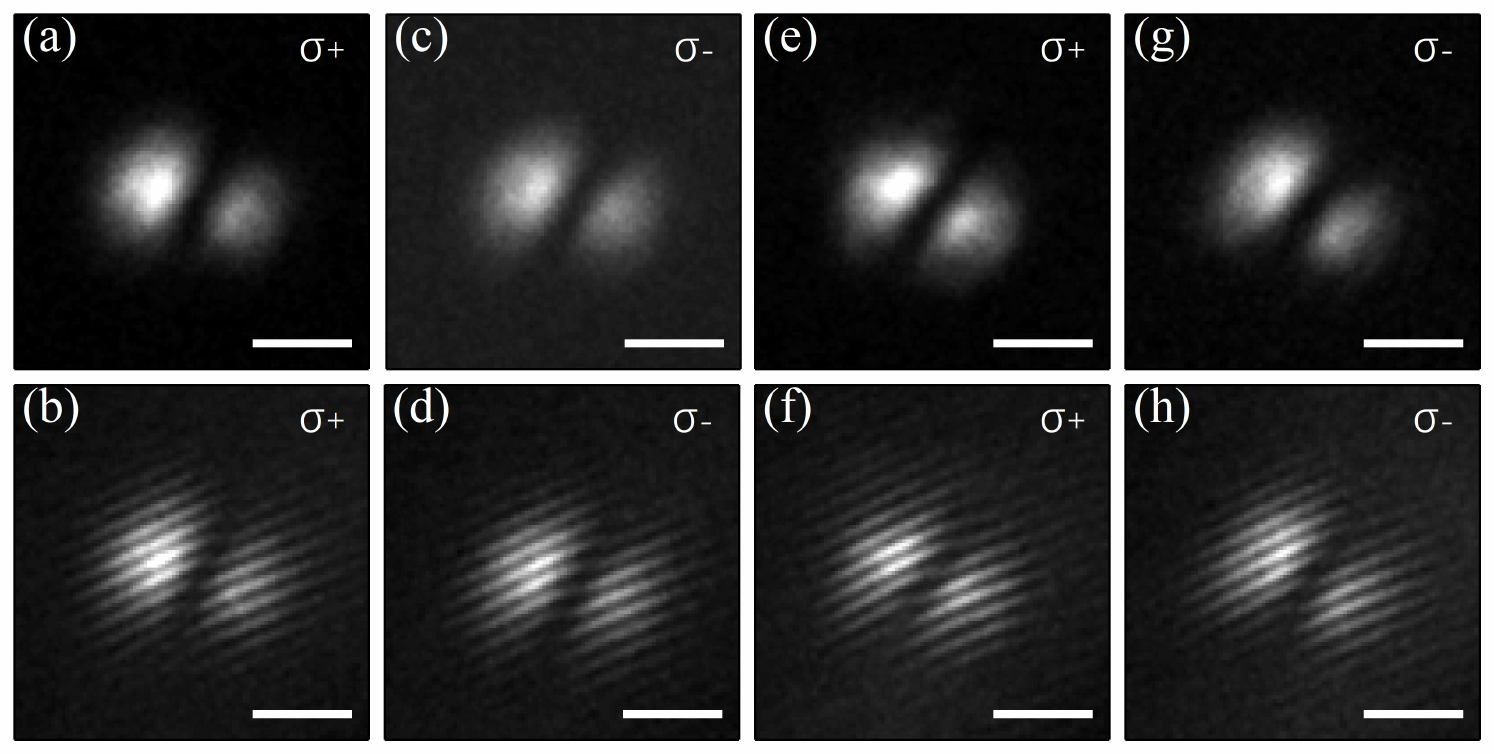}
		\caption{\textbf{Polarization dependence of polariton condensates outside the RDSOC regime.} (a, b) Real space image and interferogram of the $\sigma+$ polarized component of the polaritons at 5 V. (c, d) Real space image and interferogram of the $\sigma-$ polarized component of the polaritons at 5 V. (e-h) Same as (a-d) but for an external voltage of 6 V. The scale bars: $2\,\mu$m. }
	\end{figure}

	To summarize, we observe spin-orbit locked polariton vortex pairs in a LC microcavity. The spin-orbit locked polariton vortices originate from the TE-TM splitting. They become observable due to the separation of the condensate resulting from the non-Hermiticity induced nonreciprocal transportation behavior within the RDSOC bands for different spin component. Our results illustrate an important electrical method for the manipulation of the spin-orbit locked topological defects based on a quantum fluid of light at room temperature, and pave the way to investigate vortex pair underlying quantum statistics \cite{theo spatial} using macroscopically coherent states of hybrid light-matter particles at room temperature. These observations are of interest also for other areas such as cold atom gases and 2D electron gases in which the targeted design of spin-orbit locked states is often not accessible or much more complex to achieve.

	\begin{acknowledgments}
		TG acknowledges support from the National Natural Science Foundation of China (grant No. $12174285$). The Paderborn group acknowledges support by the Deutsche Forschungsgemeinschaft (German Research Foundation) through the transregional collaborative research center TRR142/3-2022 (231447078, project A04) and by Paderborn Center for Parallel Computing, PC$^2$.
	\end{acknowledgments}


\end{document}